# Magmatic volatiles to assess permeable volcano-tectonic structures in the Los Humeros geothermal field, Mexico

Anna Jentsch[a,b,*], Egbert Jolie[a], David G Jones[c,1], Helen Taylor-Curran[c], Loïc Peiffer[d], Martin Zimmer[a], Bob Lister[c],

[a] GFZ German Research Centre For Geosciences, Telegrafenberg, 14473 Potsdam, Germany

[b] Institute of Geosciences, University of Potsdam, 14476 Potsdam, Germany

[c] BGS British Geological Survey, Nicker Hill, Keyworth, Nottingham NG12 5GG, United Kingdom

[d] CICESE Centro de Investigación Científica y de Educación Superior de Ensenada, Carr Tijuana-Ensenada 3918, 22860 Ensenada, Baja California, Mexico

*Corresponding author: Anna Jentsch, GFZ German Research Centre For Geosciences, Telegrafenberg, 14473 Potsdam, Germany (ajentsch@gfz-potsdam.de)

**Abstract**

Magmatic volatiles can be considered as the surface fingerprint of active volcanic systems, both during periods of quiescent and eruptive volcanic activity. The spatial variability of gas emissions at Earth's surface is a proxy for structural discontinuities in the subsurface of volcanic systems. We conducted extensive and regular spaced soil gas surveys within the Los Humeros geothermal field to improve the understanding of the structural control on fluid flow. Surveys at different scales were performed with the aim to i) identify areas of increased gas emissions (reservoir scale), ii) their relation to (un)known volcano-tectonic structures (fault scale) favoring fluid flow, and iii) determine the origin of gas emissions. Herein, we show results from a $CO_2$ efflux scouting survey, which was performed across the main geothermal production zone (6 km x 4 km) together with soil temperature measurements. We identified five areas with increased $CO_2$ emissions, where further sampling was performed with denser sampling grids to understand the fault zone architecture and local variations in gas emissions. $CO_2$ efflux values range from below detection limit of the device to 1,464 g m$^{-2}$ d$^{-1}$ with a total output of 87 t d$^{-1}$ across an area of 13.7 km$^2$. Furthermore, $\delta^{13}C_{CO2}$ and $^3He/^4He$ analyses complemented the dataset in order to assess the origin of soil gases. Carbon isotopic data cover a broad spectrum from biogenous to endogenous sources. Determined $^3He/^4He$ ratios indicate a mantle component in the samples of up to 65 % being most evident in the northwestern and southwestern part of the study area. We show that a systematic sampling approach on reservoir scale is necessary for the identification and assessment of major permeable fault segments. The combined processing of $CO_2$ efflux and $\delta^{13}C_{CO2}$ facilitated the detection of permeable structural segments with a connection to the deep, high-temperature geothermal reservoir, also in areas with low to intermediate $CO_2$ emissions. The results of this study complement existing geophysical datasets and define further promising areas for future exploration activities in the north- and southwestern sector of the production field.

---

[1] 30 North Road, West Bridgford, Nottingham, NG2 7NH, UK

## 1. Introduction

Volcanic-hosted geothermal systems comprise a vast amount of thermal energy with the potential to reach supercritical conditions (T > 374°C and P > 221 bar for pure water) nearby magmatic intrusions (Scott et al., 2015). The identification, characterization, assessment, and development of exploitation concepts for the Los Humeros superhot geothermal resource is the focus of the Mexican-European collaborative project GEMex (Jolie et al., 2018). The successful utilization of the superhot reservoir is expected to increase the overall productivity of the field. However, technical challenges associated with reservoir fluids of aggressive physicochemical characteristics, drilling into high-temperature zones > 350°C, and insufficient formation permeability need to be overcome (Gutiérrez-Negrín and Izquierdo-Montalvo, 2010). The structural control of permeability and fluid flow in volcanic-geothermal reservoirs like Los Humeros has a substantial influence on the geothermal potential of the system (Jafari and Babadagli, 2011). Hydrothermal fluid flow occurs preferentially along major subsurface discontinuities (Curewitz and Karson, 1997; Rossetti et al., 2011; Rowland and Sibson, 2004; Jolie et al., 2016, 2019).

Previous studies at Los Humeros focused already on the structural architecture of the large silicic caldera (Campos-Enriquez and Arredondo-Fragoso, 1992), but thick layers of post caldera volcanic material are expected to cover many faults and fractures (Norini et al., 2015). Fluid migration is mainly controlled by a pronounced fault network (Arellano et al., 2003; Norini et al., 2015; Peiffer et al., 2018), while petrophysical analyses of samples from the reservoir units indicate low to medium permeability (Weydt et al., 2018). Geophysical surveys (e.g., resistivity, gravity, seismicity) are commonly used in geothermal exploration, but cannot resolve single permeability structures rather than wide zones. More importantly, resistivity imaging is limited in its differentiation between active and past hydrothermal activity. In this context, soil gas surveys substantially complement established geophysical exploration techniques by indicating recent volcanic-geothermal activity. Spatial variations of gas emissions and their isotopic composition can be related to permeable segments of the structural framework in a volcanic complex, and provide clear evidence of an active geothermal system with hydrothermal fluid circulation. Therefore, we performed an extensive, regularly-spaced $CO_2$ efflux survey at Los Humeros Volcanic Complex (LHVC) with the aim to identify and quantify areas of increased gas emissions as indicator of permeable pathways connecting the deep volcanic-geothermal system with Earth's surface.

$CO_2$ is one of the most abundant gases in volcanic-geothermal systems as it exsolves from magma at greater depth (Edmonds and Wallace, 2017). $CO_2$ emissions can be measured in-situ by the accumulation chamber technique (Parkinson, 1981) with a sampling time of only 2-3 min per site. Additionally, we sampled from selected sites for carbon and helium isotope analyses to determine the origin of fluids and assess transport processes of $CO_2$ and He. Helium isotopes ($^3$He/$^4$He) are excellent tracers to differentiate between crustal or mantle derived fluids, since both isotopes are stable, chemically inert and insoluble in water (Ozima & Podosek, 2002). $^3$He was trapped in the mantle during Earth accretion, while $^4$He is a decay product of the natural occurring radionuclides ($^{238}$U, $^{235}$U, $^{232}$Th) and present in the crust and atmosphere (Karlstrom et al., 2013; Notsu et al., 2001; Ozima & Podosek, 2002). Based on results from the large-scale $CO_2$ efflux scouting survey, multiple domains with increased $CO_2$ emissions were identified for a more specific assessment by smaller grids. The domain-based approach was applied to gain a high-resolution picture across areas with increased gas emissions, to identify segments of highest permeability, and to understand the heterogeneity of gas emission over short distances, especially along fault damage zones (Caine et al., 1996; Jolie et al., 2016).

## 2. Geology

The Trans-Mexican Volcanic Belt (TMVB) is a 1,000 km long Neogene volcanic arc resulting from the subduction of the two oceanic plates, Cocos and Rivera, under the North American Plate (Pérez-Campos et al., 2008; Fig. 1a). Three of the five geothermal production fields used for power generation in Mexico are located within the TMVB (Gutiérrez-Negrín, 2019). The LHVC is the largest, silicic caldera complex in the eastern part of the TMVB (Carrasco-Núñez et al., 2017b; Fig. 1a) and hosts a high-temperature geothermal system, which is currently utilized by a geothermal power plant with an installed capacity of 95 MWe (Gutiérrez-Negrín, 2019).

## 2.1 Geological history

The Los Humeros caldera has a complex geological and tectonic evolution studied in detail during the early stages of exploration by Ferriz and Mahood (1984), who dated the beginning of the caldera forming volcanism to 460 ± 40 ka lasting until 50 ± 20 ka ago. Recently, Carrasco-Núñez et al. (2018) determined a much shorter time frame for the caldera forming stage (164 ± 4.2 ka until 69 ± 16 ka ago) and concluded that the existence of such a high enthalpy geothermal system must be related to a young heat source. The following stratigraphic description refers mainly to the work of Carrasco-Núñez et al. (2017a,b; 2018) and references therein. The pre-, syn-, and post-caldera evolution is marked by alternating episodes of explosive and effusive eruptions producing a large range of volcanic rocks from different sources and depths (Lucci et al., 2019). The basement of the LHVC is comprised of a thick layer of Paleozoic granites and Mesozoic limestones with major mafic and silicic intrusion overlain by andesitic and basaltic lavas (Teziutlan Formation; 1.44 – 2.65 Ma) (Carrasco-Núñez et al., 2017a). The onset of the caldera forming period was dominated by an immense, explosive eruption 164 ± 4.2 ka ago causing the irregular shaped Los Humeros caldera with a diameter of approximately 20 km (Carrasco-Núñez et al., 2018; Fig.1b). Followed by a period of plinian eruptions producing different pumice fallouts known as the Faby Tuff, a second major caldera forming event formed the Los Potreros caldera 69 ± 16 ka ago (Carrasco-Núñez et al., 2017a). This semicircular caldera with a diameter of around 10 km hosts the present-day active geothermal system. During the post-caldera episode, volcanic activity inside the caldera moved from the central part of the geothermal field to the northern and southern areas where volcanic activity mainly occurred along ring faults (monogenetic eruptive centers). Another significant eruption inside the Los Humeros caldera occurred 7.3 ± 0.1 ka ago and left the oval shaped Xalapazco crater (approx. 1.7 km in diameter, Fig.1b) in the south of the complex (Carrasco-Núñez et al., 2018; Ferriz and Mahood, 1984; Willcox, 2011). The youngest activity occurred 2.8 ± 0.03 ka ago, which produced a trachytic lava flow in the vicinity of the SW caldera rim (Carrasco-Núñez et al., 2017a).

## 2.2 Structural evolution

The Los Humeros geothermal reservoir, hosted by the pre-caldera andesites, is characterized by medium to low matrix permeability (Gutiérrez-Negrín and Izquierdo-Montalvo, 2010). Hence, main geothermal fluid flow can only be controlled by a dense fault/fracture network resulting from different periods of volcano-tectonic activity (Norini et al., 2019). Early structural analysis by Campos-Enriquez and Garduño-Monroy (1987) concentrated on the regional tectonic regime and its impact on the local fault system within the LHVC, which revealed two dominant regional structural systems of NW-SE and NE-SW orientation. Further extensive structural fieldwork by Norini et al. (2015; 2019) across the LHVC and surrounding areas confirmed the influence of the two dominant regional structural systems on faults inside the caldera. The initial regional fault system evolved from a NE-SW compressive orogenic phase resulting in the Mexican Fault and Thrust Belt (MFTB) leaving NW-SE/NNW-SSE oriented folds and thrust faults in the sedimentary basement ($S_{Hmax}$ = NE-SW; $S_{hmin}$ = NW-SE). These basement structures favored the later development of the main NW-SE/NNW-SSE normal fault system (e.g., Los Humeros and Maztaloya fault), which developed during caldera collapse and post caldera volcanic activity under an extensional regime (NE-SW). The second fault system is related to the TMVB evolution, which evolved under a regional NE-SW extensional stress regime ($S_{Hmax}$ = NW-SE; $S_{hmin}$ = NE-SW) evolving NE-SW striking normal faults. Faults like Pedernal or Cueva Ahumada (Fig. 1c) are local volcanotectonic faults, which were generated after the Los Potreros caldera collapse but their morphology may be a heritage of TMVB regional faults as they follow a NE-SW strike. The present-day faulting activity is influenced by an extensional (regional) and radial (local) stress field, which relates to the shallow magmatic/hydrothermal system with recent normal and reverse faulting (Fig. 1c; Norini et al., 2019).

Past and recent hydrothermal fluid flow is indicated along known deep-rooted faults (e.g., La Cuesta, Los Humeros, Arroyo Grande) connected with the geothermal reservoir (secondary permeability) (Norini et al., 2019). Deep-rooted regional faults within the basement of the caldera are indicated by water isotopic analyses of fluid samples from springs and wells in and around the LHVC, suggesting a regional recharge of the geothermal system (Lelli and Villanueva Alfaro Cuevas, 2019).

**Figure 1**

### 3. Methods

**3.1 Sampling approach**

Systematic soil gas surveys in volcanic and geothermal areas require the development of site-specific sampling concepts. The decision of whether or not an area is investigated by transects or (ir)regular grids depends on the objective of the survey, time, accessibility to the area, and site-specific characteristics such as the presence of fumaroles, alteration, known structures and other factors (Fridriksson et al., 2006; Hernández et al., 2012; Parks et al., 2013; Rodríguez et al., 2015). Specific areas with obvious gas emissions are often sampled with a higher density of measurements for a more accurate detection of variations in soil gas emissions and delineation, whereas areas with less obvious gas emissions are often not taken into account (Werner et al., 2000; Chatterjee et al., 2019) despite the chance of missing permeable structures (hidden faults). Multiple studies showed the advantage of regular spaced surveys (e.g., Werner et al., 2000; Jolie et al., 2019). Systematic grid sampling is an effective and unbiased concept to collect spatially correlated data over a large area (geothermal reservoir scale) whilst avoiding interpolation artefacts (Isaaks and Srivastava, 1989) and allowing a detailed estimation of the total $CO_2$ output (combined advective and diffuse degassing) (Lee et al., 2016).

For the $CO_2$ scouting survey (2017), we developed a regular sampling network for an area of 6 km × 4 km adapted specifically to the structural setting. The network consists of 2,700 $CO_2$ efflux-sampling points with the purpose to detect spatial variability of soil gas emissions across major parts of the geothermal production field. The sampling grid is defined by 25 m × 200 m spacing with the small point spacing oriented perpendicular to the major fault strike (NNW-SSE) within the caldera (Table 1, Fig. 3a). $CO_2$ efflux was measured in-situ (60–120 s) by means of the accumulation chamber method (Chiodini et al., 1998). Due to expected anthropogenic disturbances no $CO_2$ efflux measurements were performed in Humeros village.

Based on the results of the $CO_2$ efflux scouting survey, five main areas have been identified with increased $CO_2$ emissions. These areas (Fig. 3a) were investigated at higher resolution in 2018 (Table 2) by the accumulation chamber method. The domain-based approach facilitated a more detailed estimation of $CO_2$ output for the investigated areas.

**Table 1 and Table 2**

**3.2 $CO_2$ efflux and soil temperature**

$CO_2$ efflux measurements were performed with portable LICOR LI-820 infrared gas analyzers being connected through a closed loop to an accumulation chamber Type A (West Systems, 2019). The LI-820 is a non-dispersive, infrared (NDIR) gas analyzer based upon a single path, dual wavelength detection system. The measuring range is between 0 and 20,000 ppm with a maximum gas flow rate of 1 l min$^{-1}$ (West Systems, 2019). A detailed description of the accumulation chamber method is defined by Chiodini et al. (1998). All measurements and sampling took place under dry weather conditions. Soil temperatures were measured at 50 cm depth with a Greisinger GMH 285-BNC thermometer (accuracy ± 0.1 °C) coupled to a 620 mm long stainless steel probe. For the domain-based surveys a Hanna HI-93510® thermistor thermometer (accuracy ± 0.4 °C) fitted to a 500 mm temperature probe was used.

The statistical evaluation of $CO_2$ efflux data was performed by the graphical statistical analysis (GSA) introduced by Sinclair (1974; Fig. 2). This method is commonly used to separate large datasets by plotting the logarithmic $CO_2$ efflux values against the cumulative frequency and identifying major inflection points as a key indicator of different populations. By identifying multiple log-normal populations different gas sources and/or transport mechanisms can be inferred. A detailed description of the GSA method was compiled by Chiodini et al. (1998, 2008). Both datasets (scouting and domain-based $CO_2$ surveys) have been merged for statistical analyses.

The interpolated maps of $CO_2$ efflux and soil temperature were generated by means of sequential Gaussian simulations (sGs) using ESRI ArcGIS® 10.5 software. The sGs algorithm was introduced by Deutsch and Journel (1998), who emphasize that sGs respects original data without smoothing extreme values in order to preserve spatial variations. In total, 100 realizations were performed for the sGs based on a simple kriging model. The sGs procedure requires a gaussian distribution of data. Therefore, all data were normal score transformed and declustered to correct data distribution before the generation of omnidirectional variograms. All variograms and related model parameters can be found in the supplementary material. Interpolated maps have been computed up to 97 g m$^{-2}$ d$^{-1}$, since merely 3% of the data show values ≥ 97 g m$^{-2}$ d$^{-1}$. The strong variation of effluxes over short distances respects high values more than low resulting in a different appearance of anomalies. Through the comparison of the different interpolations (map including all values versus map including only values up to 97 g m$^{-2}$ d$^{-1}$), we assessed that the exclusion of values ≥ 97 g m$^{-2}$ d$^{-1}$ in our interpolation method emphasizes the spatial extent of lower degassing areas with hydrothermal signatures, which play a major role regarding structural related degassing. This way of data presentation still respects all information since values above 97 g m$^{-2}$ d$^{-1}$ are illustrated as graduated black triangles on maps.

**Figure 2**

### 3.3 $\delta^{13}C$-$CO_2$

94 soil gas samples were collected for $\delta^{13}C_{CO2}$ isotopic analyses (Table 1 & 2) from areas of low, intermediate and high $CO_2$ efflux to constrain the origin of $CO_2$ emissions by their isotopic signatures (biogenous, endogenous). A hollow metal probe was installed in 1 m depth for gas sampling. To avoid any contamination with ambient air, the metal probe was flushed five times with a 60 ml syringe before collecting a gas sample. The sample was injected into an evacuated 12 ml glass vial with a pierceable septum. Analyses were performed with a GC-C-IRMS system consisted of a GC (6890N, Agilent Technology, USA) connected to a GC-C/TC III combustion device coupled via open split to a MAT 253 mass spectrometer (ThermoFisher Scientific, Germany) under continuous flow. The quality of the carbon isotope measurements was checked by direct injection a $CO_2$-reference gas with known $\delta^{13}C$ into the mass spectrometer during each run as well as by daily measurements of n-alkane gas standard (n-C1 to n-C6). The standard deviation is < 0.5 ‰. Herein, all results are reported in the standard notation δ as per mille (‰) deviations relative to the PDB (Pee Dee Belemnite) standard.

### 3.4 $^3$He/$^4$He ratios

To further constrain the origin of gas emissions, six helium samples were taken from selected sites to determine the $^3$He/$^4$He ratios. Samples were taken on steam vents to reduce the contribution of $^4$He from ambient air by placing a funnel on the outlet of the steam vent, which was connected via a flexible tube to a 40 cm copper pipe. If steam flow was insufficient, gas was aspirated with a hand pump for about two minutes. Metal clamps were fixed to each end of the copper tube and tightly closed for a safe storage of the gas. The samples were analyzed with a VG-5400 mass spectrometer. Results were normalized to the air ratio ($R/R_A = 1.386 \times 10^{-6}$).

## 4. Results

A summary of the different parameters is presented in Table 1 & 2. $CO_2$ efflux values from the two field campaigns show a wide range and vary from below detection limit to 1,464.2 g m$^{-2}$ d$^{-1}$. By applying the GSA method two inflection points have been identified at the 90 % and 99.8 % cumulative percentile, which separate the dataset into three populations with a gently sloping central segment (Fig. 2). Sinclair (1974) explains this kind of shape as characteristic for an overlap of two populations (A+B). Population A corresponds to low background efflux values (0.1 – 20.7 g m$^{-2}$ d$^{-1}$). Population B is a mixed population with a wide range of values (20.7 – 640.8 g m$^{-2}$ d$^{-1}$), and Population C includes the highest measured values (670.8 – 1,464.2 g m$^{-2}$ d$^{-1}$) in the study area. Statistical parameters obtained by the GSA method are reported in Table 3. The low fraction of high degassing values is related to our sampling approach, which follows a regular sampling grid rather than mapping exclusively areas of high $CO_2$ emissions with only a few sites of background values. In consequence, the domain-based survey in 2018 resulted in more data points assigned to population C as the focus was on areas of increased $CO_2$ emissions.

**Table 3**

### 4.1 $CO_2$ efflux scouting survey and soil temperatures

Figure 3a shows the result of sequential Gaussian simulations (mean out of 100 simulations) from the $CO_2$ efflux scouting survey in 2017. Bold capital letters in Figure 3a refer to results explained in the following paragraph. Zones with gas emissions exceeding the background threshold (> 20.7 g m$^{-2}$ d$^{-1}$) occur across the entire study area (Area A to G). Areas A, C, D, and E define a well pronounced NNW-SSE corridor (3 km × 1.5 km) of increased gas emissions > 100 g m$^{-2}$ d$^{-1}$. The maximum $CO_2$ efflux was measured in Area E with 839 g m$^{-2}$ d$^{-1}$ (Fig. 3a). La Cuesta fault in Area A and Los Humeros fault in Area D show increased $CO_2$ degassing in their foot- and hanging walls (< 200.9 g m$^{-2}$ d$^{-1}$). Areas A to E are characterized by major $CO_2$ emissions, which occur together with geothermal surface manifestations, such as weak steam vents, sparse solfatara, and hot ground with temperatures up to 91°C, and argillic alteration (Gutiérrez-Negrín and Izquierdo-Montalvo, 2010). However, areas of increased $CO_2$ emissions are not limited to these features. There are no significant $CO_2$ emissions or geothermal surface expressions in the south of the study area (Area H, Fig. 3a). Almost no gas emissions have been detected in Area I, which is covered by a young basalt flow (Carrasco-Núñez et al., 2017a).

Soil temperatures at 50 cm depth vary from 5.7 °C to 91.3 °C (Table 1). The histogram (Fig. 4) shows a right skewed frequency distribution for soil temperatures illustrating that the majority of values were low, while the probability plot indicates a prominent inflection point at 22 °C. Increased soil temperatures (> 22 °C) occur in or close to areas of increased degassing (Area A-E; Fig. 5a). In addition,

Area J and K show increased temperatures of 41 °C and 26 °C, respectively (Fig. 5a). The maximum measured soil temperature of 91.3 °C occurred in Area C, where active solfatara is located.

**Figure 3**

**Figure 4**

**Figure 5**

### 4.2 Domain-based $CO_2$ efflux and soil temperature survey

Figure 3b shows diffuse $CO_2$ emissions of four selected domains out of five. All values below 29 g m$^{-2}$ d$^{-1}$ are masked. Highest $CO_2$ efflux values of 1,285.5 g m$^{-2}$ d$^{-1}$ and 1,464.2 g m$^{-2}$ d$^{-1}$ were measured in Area E and A, respectively. All three areas (A, D, E) are characterized by a general NNW-SSE orientation of increased $CO_2$ emissions. Area D is an excellent example of highly variable $CO_2$ emissions over short distances resolved by a grid spacing of 10 m × 10 m. In comparison to the large-scale scouting survey, where increased $CO_2$ degassing appears less variable, the domain-based surveys improved the spatial resolution in these areas. Nevertheless, both sampling concepts provide similar results for the occurrence of major diffuse $CO_2$ emissions despite different grid spacing. During domain-based sampling, soil temperatures were solely measured in Area B and E (Fig. 5b & Table 2). A maximum temperature of 52.9 °C was measured in Area B coinciding with increased $CO_2$ emissions. In Area E well defined cluster have been identified with soil temperatures up to 35 °C.

### 4.3 Isotopic analyses

Carbon isotopic values range from -23.2‰ to -1.2‰ ± 0.07‰ (Table 1 & 2), covering a broad spectrum of sources (Fig. 3a & b). Steam vents were selected for $^3$He/$^4$He ratio analyses (Fig. 5a). $^3$He/$^4$He ratios deviate for all samples from the air helium isotopic ratio ($R_A$ = 1.386 × 10$^{-6}$) and confirm a mantle contribution. The measured ratios range from 2.31 ± 0.58 to 4.88 ± 0.99. Highest $^3$He/$^4$He ratios were identified in Area A and E, coinciding with maximum measured $CO_2$ emissions.

### 4.4 $CO_2$ output estimations

Based on the sequential Gaussian simulation maps for the investigated areas, a $CO_2$ output estimation was computed. Each $CO_2$ degassing rate is the sum of the product of each grid cell by the cell surface. The total $CO_2$ output (biogenous plus endogenous) is estimated to be 87.6 t d$^{-1}$, whereas the $CO_2$ output calculated above the biogenic threshold only accounts for 26.1 t d$^{-1}$. A detailed summary of results from all areas is given in Table 4, which also includes $CO_2$ degassing rates computed by Peiffer et al. (2018). It is noteworthy that their results are based on random and dense sampling points concentrating on known high degassing areas in Los Humeros. For comparison, we included further $CO_2$ output estimations from other volcanic-geothermal systems worldwide (Table 4). A comprehensive data compilation of $CO_2$ output estimations from various active, dormant, and inactive volcanic regions worldwide is summarized in a study of Kis et al. (2017).

**Table 4**

## 5. Discussion

### 5.1 $CO_2$ efflux and $\delta^{13}C_{CO2}$

#### 5.1.1 Population A

The majority of measured $CO_2$ efflux values belong to population A with a mean of 6.6 g m$^{-2}$ d$^{-1}$ (Table 3). The inflection point separating the background population from the remaining values is defined at 20.7 g m$^{-2}$ d$^{-1}$ (Fig. 2). Such low efflux values typically originate from biogenic sources (i.e. plant and microbial respiration and organic decomposition; see also Werner et al., 2000; Cardellini, 2003; Peiffer et al., 2014; Hutchison et al., 2015; Jolie et al., 2019). In fact, maximum $CO_2$ efflux values reported for biogenic sources generally range from below detection limit of the device up to a few tens of grams per square meter and day in highly vegetated areas, but never exceed 100 g m$^{-2}$ d$^{-1}$ (Chiodini et al., 2007; Raich and Tufekcioglu, 2000; Widén and Majdi, 2011). Peiffer et al. (2018) report similar results for areas without any apparent geothermal manifestations in the Los Humeros caldera with a mean of 7.4 g m$^{-2}$ d$^{-1}$. The overall vegetation within the survey area is very sparse and dominated by shrubs and dry grasses growing on loose soil consisting of pumice tuff mixed with organic material. However, to some extend the northern part is used for maize cultivation, whereas the south is partly forested by pine trees. Additionally, most $\delta^{13}C_{CO2}$ samples acquired in this study confirm this observation by heavier $\delta^{13}C_{CO2}$ values < -10‰. Figure 6 illustrates the relation between $CO_2$ efflux and corresponding carbon isotopic composition in context to previously published carbon isotopic data from Los Humeros (Truesdell and Quijano 1988; González-Partida et al., 1993; Portugal et al. 1994; Peiffer et al., 2018; Richard et al., 2019).

We have identified seven locations with $CO_2$ efflux values lower than 20.7 g m$^{-2}$ d$^{-1}$, but with a $\delta^{13}C_{CO2}$ isotopic composition ranging from -3.9‰ to -7‰ (Area A, D, E) indicating a clear contribution of hydrothermal and mantle-derived $CO_2$ (Fig. 6). The identified locations are up to 200 m distant to areas with increased gas emissions, which indicates that the actual dimension of geothermally active areas is larger than expected. This finding agrees with Chiodini et al. (2008), who estimated a larger size of a diffuse degassing structure at Solfatara of Pozzuoli by using the biogenic threshold resulting from $\delta^{13}C_{CO2}$ analysis of respective $CO_2$ effluxes. Following the approach of Chiodini et al. (2008) our carbon isotopic dataset is separated in three populations (Table 5; Fig. 3 in the supplementary data). The range for biogenic $CO_2$ effluxes determined by the separation of carbon isotopes shows a similar range based on the statistical separation of the $CO_2$ efflux dataset (Population A). Whereas the range of $CO_2$ effluxes resulting from $\delta^{13}C_{CO2}$ analysis in the mixed and hydrothermal groups, show much lower minimum values (1 g m$^{-2}$ d$^{-1}$) than those identified by the statistical separation of the $CO_2$ efflux dataset in population B and C (Table 5). These low emissions occurring in the mixed and hydrothermal group are likely caused by advection at low rates, similar to biogenic emission rates, due to low permeability of soil/rocks and/or low-pressure gradients from the reservoir to the surface. Diffusion of hydrothermal $CO_2$ can also explain low emissions, if $CO_2$ concentration gradients are present within the soil system (Peiffer et al., 2014). The combined analysis of $CO_2$ efflux and carbon isotopes for a differentiation of carbon sources is a powerful approach to estimate the actual extent of geothermally active areas.

Almost no gas emissions have been detected in Area I (Fig. 3a), where a young and compact basalt flow (3.8 ka) covers the area (Carrasco-Núñez et al., 2017a). The presence of permeable structures with increased gas emissions could not be proven, but cannot be ruled out as gas efflux measurements are hardly possible to perform on compact bedrock. This is supported by a 10 m topographic scarp displacing the lava flow in Area I identified by Norini et al., (2019) which is clearly visible on the high-resolution DEM. Thermal imagery from drones could complement gas emission data in this area, as well as $CO_2$ concentration measurements with a TDL system (Tunable Diode Laser), which helps to discriminate ambient from elevated $CO_2$ concentrations possibly linked to hidden structures as also shown by Mazot et al. (2019).

### 5.1.2 Population B

The relatively large range of $CO_2$ emissions in population B (20.7 - 640.8 g m$^{-2}$ d$^{-1}$) results from multiple $CO_2$ sources and transport mechanisms. Corresponding $\delta^{13}C_{CO2}$ values range from -18 ‰ up to -1.2 ‰. All $CO_2$ efflux values between 20.7 and 29 g m$^{-2}$ d$^{-1}$ are characterized by biogenic $\delta^{13}C_{CO2}$ values. The mixed and hydrothermal $\delta^{13}C_{CO2}$ groups have corresponding $CO_2$ efflux values from 20.7 – 76.3 g m$^{-2}$ d$^{-1}$ and 22.2 – 640.8 g m$^{-2}$ d$^{-1}$, respectively (Fig. 6). This suggests that $CO_2$ emissions originate from biogenic sources, volcanic degassing, and contributions from the sedimentary basement driven by diffusive and advective transport mechanisms.

### 5.1.3 Population C

Population C (640.8 g m$^{-2}$ d$^{-1}$ – 1,468 g m$^{-2}$ d$^{-1}$) is characterized by the heaviest $\delta^{13}C_{CO2}$ values (-3.9‰ to -3‰), indicating the contribution of the volcanic-hydrothermal system. Similar $\delta^{13}C_{CO2}$ values are reported by Peiffer et al. (2018) for areas with $CO_2$ emissions > 675 g m$^{-2}$ d$^{-1}$. Separation between Population B and C is the result of solely advective gas transport, also supported by a small deviation of corresponding $\delta^{13}C_{CO2}$ values due to high $CO_2$ efflux (Camarada et al., 2007).

**Table 5**

### 5.2 Carbon isotopes - Origin and processes influencing their variability

### 5.2.1 Sources

Various studies at Los Humeros present comprehensive datasets on carbon isotope data from well fluids, calcite scales analyzed from well cuttings, and limestone from analogue outcrops of the basement (Truesdell and Quijano 1988; González-Partida et al., 1993; Portugal et al., 1994). Richard et al. (2019) sampled production steam from 20 wells and found $\delta^{13}C_{CO2}$ values ranging from -5.3 to -2.2‰. Peiffer et al. (2018) conducted three domain-based $CO_2$ efflux surveys at Los Humeros complemented by carbon isotopic analyses of samples from areas with increased $CO_2$ emissions ($\delta^{13}C_{CO2}$ between -7.8 and -2.7‰). We assigned measured $\delta^{13}C_{CO2}$ values to three groups according to characteristic $\delta^{13}C_{CO2}$ ranges for biogenic, mixed, and hydrothermal gases from literature mentioned above. The biogenic group (-23 to -10‰) is consistent with derivation of carbon from C4 (-16 to -9‰, e.g. maize; from Hoefs, 2009) and C3 plants (-33 to -23‰, e.g., high latitude grasses; from Kohn, 2010; Sharp, 2017), both present at Los Humeros. The mixed group ranges from -10 to -5‰, indicating a combination of different sources such as atmospheric (-8.3 to -7.2‰; from Ciais et al., 1995) and mantle derived $CO_2$ (-6 ± 2‰ canonical mantle value; from Marty and Zimmermann, 1999) with a small overlap to C4 plants. The hydrothermal group includes $\delta^{13}C_{CO2}$ values from -5 to -1.2‰, comprising values from the hydrothermal system (-3.3 ‰; average value of production steam from LHVC; from Richard et al., 2019) as well as sedimentary decarbonation resulting from the underlying pre-volcanic limestone basement (0.7 to -3.9‰ from calcite in well-cuttings and 0.3 to -0.8‰ from limestone basement; González-Partida et al., 1993). Following the approach of Chiodini et al. (2008), who sampled carbon isotopes from fumaroles to define a characteristic $\delta^{13}C_{CO2}$ value representative for a 'pure' magmatic fluid, we calculated a mean value for $\delta^{13}C_{CO2}$ of -3.6‰ from six anomalous degassing sites (in Area A to E) at Los Humeros. This value represents the average hydrothermal component of $CO_2$ from the reservoir, comparable to the mean

$\delta^{13}C_{CO2}$ of -3.3 from production steam determined by Richard et al., (2019). Thus, $\delta^{13}C_{CO2}$ values with hydrothermal indications derived from surficial gas emissions provide a vital tool for the identification of areas with a connection to the deep reservoir. Our results prove that reliable reservoir information can already be obtained by surface exploration methods without the necessity of drilling costly wells.

**Figure 6**

### 5.2.2 Processes

There are different processes causing a range of endogenous $\delta^{13}C_{CO2}$ values, which is subduction of the Rivera and Cocos plate under the North-American plate (Richard et al., 2019) and isotopic fractionation processes.

During the subduction process, the lithospheric mantle is mixing with fractions of the mantle wedge and carbon from different sources seems to reach the upper crust (Richard et al., 2019). Peiffer et al. (2018) and Mason et al. (2017) discuss the spread of $\delta^{13}C_{CO2}$ values by different carbon isotopic fractionation processes and propose following possibilities: (i) diffusion, (ii) partial dissolution of $CO_2$ into groundwater (both processes are present in the shallow subsurface), (iii) carbon remobilization during subduction through metamorphic decarbonation, (iv) dissolution of meta-carbonates in the basement accompanied by crustal carbon assimilation as a result of magma interaction with the crust, and (v) boiling causing calcite precipitation. Camarda et al. (2007) explain that it is crucial at which depths $\delta^{13}C_{CO2}$ samples are taken. Fractionation in the uppermost soil layers (< 1 m) due to diffusion can be neglected for our study due to sufficient sampling depth (1 m) of carbon isotopes. Verma (2000) reported results from geochemical and radiogenic isotope analyses of basaltic and rhyolitic volcanic rocks, which evidence the assimilation of crustal material in the upper mantle during magma formation at Los Humeros. Independent of a mantle, hydrothermal, or limestone carbon isotopic signature, all values evidence a permeable connection to the high-temperature geothermal reservoir.

### 5.3 Origin of helium

Air-normalized helium isotopic ratios range from 2.31 ± 0.58 to 4.88 ± 0.99 $R_A$. All samples show mixing between mantle and atmosphere as illustrated by mixing curves after Sano and Wakita (1985; Fig. 7a). Mantle contribution ranges between 30 and 65%. The maximum contribution is related to strong degassing in Area A and E (Fig. 3a & 7b). Pinti et al. (2017) report a mean $^3He/^4He$ ratio from 22 production wells in Los Humeros of 7.03 ± 0.4 $R_A$ being very close to the canonical $^3He/^4He$ ratio from mid-oceanic ridge basalts (8 ± 1 $R_A$; Graham, 2008; Fig. 7a). $^3He/^4He$ ratios decrease due to the decay of uranium and thorium associated with the formation of $^4He$ within the crust. The subsurface stratigraphy in Los Humeros is dominated by alternating layers of volcanic rocks, rich in uranium (up to 4.8 ppm) and thorium (up to 19.7 ppm), especially in andesitic and rhyolitic-dacitic rocks (Carrasco-Núñez et al., 2017b). Hence, the difference between our ratios determined from steaming ground and the ones reported by Pinti et al., (2017) from fluid samples of the geothermal reservoir (between 1,600 to 3,100 m depth) results from the enrichment of $^4He$ during slow upward migration of fluids within the shallow crust along permeability structures, compared to the rapid flow rates of steam in production wells.

### 5.4 Spatial correlation of $CO_2$ efflux, $\delta^{13}C_{CO2}$ and $^3He/^4He$

In Figures 7b and 7c two diagrams of air normalized $^3He/^4He$ ratios against $CO_2$ efflux and $\delta^{13}C_{CO2}$ values are illustrated, where the positive correlation between all three investigated parameters becomes apparent. The most prominent permeable connection from Earth's surface to the high-temperature geothermal system is indicated at sampling sites along a NNW-oriented, structure-dominated corridor with deep-rooted faults, in particular in Area A and E. Area C, D1, and D2 are also characterized by a hydrothermal source of $CO_2$. $^3He/^4He$ indicates a significant mantle component between 20 and 50% (Fig. 7a). The location of these data points occurs between Area A and E within this well-defined NNW-SSE corridor suggesting that this is the main permeability zone of Los Humeros geothermal field. Area B is located northeast of the structural corridor, also with mantle components in $^3He/^4He$ and $\delta^{13}C_{CO2}$. Production wells south of Area B (Fig. 1c) may target another N-S oriented fault zone (hidden faults), east of Los Conejos fault without further evidence of anomalous degassing.

**Figure 7**

### 5.5 Thermal anomalies

Anomalous soil temperatures ($\geq 22$ °C) correlate with areas of increased degassing, except for Area K and J (Fig. 3 & 5), where only low $CO_2$ emissions (~11-21 g m$^{-2}$ d$^{-1}$) have been determined, but still stand out from the surrounding background efflux (~1-6 g m$^{-2}$ d$^{-1}$) in these areas. Increased temperatures of 26.3 °C (Area K) and 41.2 °C (Area J) occur close (< 50 m) to known faults, where hot fluids circulate at depth and heat could be transferred primarily by thermal conduction. Bloomberg et al. (2012) and Finizola et al. (2003) observed similar effects, which they explain with hydrothermal alteration preventing degassing in areas of previously high permeability. For that reason, soil temperatures can be an additional indicator for the presence of hot fluids in the shallow subsurface, even without increased gas emissions at the surface.

The gas sample from the site with the highest measured soil temperature (91.3 °C; Area C) has a $\delta^{13}C_{CO2}$ value of -2.5‰ and is surrounded by solfatara being the most obvious indicator for magmatic degassing (Francis and Oppenheimer, 2004). Espinosa-Paredes and Garcia-Gutierrez (2003) estimated static formation temperatures (SFT) for selected wells in Los Humeros (1,500 - 3,265 m). The high SFT for well H-29 (433.6 °C; depth 2,186 m, Fig. 1c), drilled into the footwall of Loma Blanca fault might explain maximum soil temperature in Area C, which is favored by the near-vertical upflow of hydrothermal fluids. Strong alteration along Loma Blanca fault supports this observation.

### 5.6 Deep-derived gases in context to fault zone architecture and geothermal production

While some of the faults in Los Humeros have prominent fault scarps, their continuation into the deep geothermal reservoir often remains unknown (Norini et al., 2015). Geophysical resistivity studies indicate zones of increased alteration as a result of intense fluid-rock interaction in the subsurface of Los Humeros reaching up to 5-6 km depth (Arzate et al., 2018; Benediktsdóttir et al., 2019). Different studies have shown that faults can act as major conduits for fluid-flow and are responsible for secondary permeability in geothermal systems (Caine et al., 1996; Rowland and Sibson, 2004; Rossetti et al., 2011; Jolie et al., 2016), but can also reduce permeability and act as a barrier due to mineral precipitation or comminution (Aben et al., 2016; Rossetti et al., 2011; Rowland and Sibson, 2004). Our study shows that $CO_2$ degassing is typically not limited to single fault planes, instead it is influenced by wide fault damage zones or multiple, interconnected faults with anisotropic and heterogeneous properties. Los Humeros fault system was influenced by different periods of volcano-tectonic activity under changing stress conditions (see chapter 2.2). Faulds and Hinz (2015) assessed different structural settings (e.g., fault steps, intersections, tips) in the Basin and Range Province favoring the formation of geothermal systems.

In the following paragraphs, we have related some of our observations to structural settings described in their study (Fig. 8a-e).

### 5.6.1 Area A – Normal fault

Degassing at La Cuesta fault follows an N-S orientation along a 400 m-long and 150 m-wide segment (Fig. 8a). La Cuesta is a normal fault promoting fluid upflow from the deep reservoir, preferentially in its footwall. This is supported by highest $^3$He/$^4$He values and hydrothermal $\delta^{13}C_{CO2}$. Since well H-35 is drilled nearly vertical and proved to be suitable as a geothermal production well, another permeable segment is indicated to the west of La Cuesta.

### 5.6.2 Area B – Hidden fault/Fault continuation

Increased soil gas emissions in Area B are limited in its extent, but might be the result of a hidden fault structure. A possible continuation of Los Conejos fault at depth towards the north cannot be excluded, unless another, hidden structural corridor, parallel to the east of Los Conejos fault is present. This could be indicated by multiple production wells targeting the geothermal reservoir at a depth of ~2800 m (Fig. 8b).

### 5.6.3 Area C – Horsetail fault termination

Degassing in Area C is observed in a 550 m E-W × 250 m N-S wide zone north of Humeros village with clustered gas emissions and soil temperatures up to 91.3°C, as well as signatures of mantle derived helium and hydrothermal carbon. The clustered gas emissions may be indicative for the existence of further structural elements, hidden beneath Humeros village. We suggest that increased fluid flow might be accommodated by a horsetail fault termination where the break up into multiple fault strands favors geothermal fluid flow (Fig. 8c). Additional $CO_2$ efflux measurements by Peiffer et al. (2018) towards the north of Area C show a continuation of increased degassing along Loma Blanca and Los Humeros fault.

### 5.6.4 Area D – Fault damage zone

The Los Humeros fault in Area D has a vertical escarpment up to 80 m with evident heterogeneous degassing in its fault damage zone (Fig. 8d & 9). Degassing along the fault core is low, typically the result of hydrothermal alteration and mineral precipitation (Wyering et al., 2014). Increased soil temperatures have been measured along the fault scarp (thermal conduction). Gas emissions along Los Humeros fault are limited to a 700 m-long NNW-SSE segment and point to the most permeable part of this large normal fault.

### 5.6.5 Area E – Fault intersection

Area E hosts the largest zone of increased gas emissions (600 m E-W × 1,000 m N-S) and hydrothermal signals ($^3$He/$^4$He, $\delta^{13}C_{CO2}$) along a well-confined corridor (Fig. 3b & 8e). Increased degassing could be related to a fault intersection between La Antigua and an unnamed thrust fault (Fig. 8e). However, also the linkage of damage zones from the unnamed thrust fault, La Antigua and maybe Los Humeros fault at depth might promote increased fluid flow in that area (Fig. 9). In some parts of Area E, we noticed superficial argillic alteration (e.g., kaolinite), a result of upward migration of geothermal fluids to the surface (Bernard et al., 2011). The orientation of the zone with anomalous degassing reflects the typical NNW-SSE orientation of faults in this structural corridor of the LHVC.

**Figure 8**

## 6. Conclusion and Outlook

We present results of a comprehensive multi-scale soil gas survey in the Los Humeros geothermal field to identify areas of increased permeability as a result of structural discontinuities in the subsurface. Our findings demonstrate that the majority of increased diffuse $CO_2$ emissions are hydrothermal/mantle derived with some contribution from metamorphic decarbonation. The combined analysis of $CO_2$ efflux and $\delta^{13}C_{CO2}$ showed that areas of background diffuse $CO_2$ emissions could still be related to fluid pathways with a connection to the deep reservoir. Helium isotopic analyses at selected locations complement results from the area-wide $CO_2$ efflux survey and prove the existence of deep-rooted faults down to the high-temperature geothermal reservoir. Independent of the source, increased gas emissions always indicate the presence of a fault-controlled fluid migration along permeable segments of fault zones. This is an important information for the definition of input parameters for dynamic models of geothermal systems. The further identification of migration pathways of hydrothermal fluids in the subsurface can be acquired by soil temperature measurements at sufficient depth even without increased gas emissions. One of the most significant areas for geothermal power generation in Los Humeros is a permeable, NNW-SSE oriented structural corridor, which is targeted by a large number of the geothermal production wells (Fig. 9) and shows the strongest mantle contribution in helium and carbon isotopes. Furthermore, we suggest another N-S oriented, structurally confined compartment, which includes Area B and is also targeted by many of the production wells (Fig. 9). A promising area of enhanced structural permeability was identified in Area E, making it a possible target for future geothermal exploration activities.

We demonstrate that the application of large-scale $CO_2$ efflux surveys with suitable sampling distance is a successful approach for geothermal exploration. Domain-based surveys with higher resolution (smaller grid spacing) improve the assessment of spatial variability of gas emissions along specific faults. Some structures are characterized by increased fluid flow along deep-rooted faults such as La Cuesta and La Antigua fault. Others may act to some extent as barriers, for example Las Papas or Los Humeros fault.

Soil gas studies are suitable exploration techniques in volcanic-geothermal systems before costly drilling operations and should become an integral part in the overall exploration strategy in new and undeveloped geothermal fields to complement established methods, e.g., geological mapping and geophysical exploration. Gas measurements in combination with bathymetric and ground temperature surveys have even proven their suitability to indicate geothermal fluid flow in unconventional geothermal settings, e.g., limnic environments (Jolie, 2019). Although, $CO_2$ emissions from anomalous degassing sites play an important role when calculating the $CO_2$ output from an area, we could show that two-thirds of the $CO_2$ emissions in Los Humeros are related to background values and should not be neglected when calculating the total $CO_2$ output (Fig. 9), which has a significant contribution to the global $CO_2$ budget.

Furthermore, the continuous monitoring of volcanic gas emissions will improve the understanding of temporal variations, their relation to seismic and/or volcanic activity, and effects of geothermal exploitation at Los Humeros, which is the focus of another publication currently in preparation. Further work should focus on variable sampling grids and integrated approaches, and confirm their reliability in different volcanic-geothermal settings.

**Figure 9**

## Acknowledgements

This paper presents results of the GEMex Project, funded by the European Union's Horizon 2020 research and innovation programme under Grant Agreement No.727550, and by the Mexican Energy Sustainability Fund CONACYT-SENER, Project2015-04-268074. The authors wish to thank the Comisión Federal de Electricidad of Mexico (CFE) for their support, and access to Los Humeros geothermal field. A special thank you to all people from Mexico, Argentina, and Germany who supported us to collect this comprehensive dataset: Esteban Silva, Sonia Vargas Pineda, Romel González, Ana Maria Dávalos Pérez, Jorge Alejandro Guevara Alday, Ruth Alfaro Cuevas Villanueva, Rojeh Khleif, Tanja Ballerstedt, Camila Espinoza, Leandra Weydt, and Adrian Lechel. Thank you to the GEMex consortium, to Gianluca Norini and Kyriaki Daskalopoulou for fruitful discussions. Thank you to the Universidad Michoacana de San Nicolás de Hidalgo for providing vehicles for the fieldwork. We would also like to thank Section 4.3: Climate Dynamics and Landscape Evolution with the laboratory for compound-specific isotope analysis and Section 3.1: Inorganic and Isotope Geochemistry with the noble gas laboratory at the German Research Center for Geoscience (GFZ) in Potsdam as well as Charles Belanger and Steve Brookes (Iso-Analytical Limited) for analysis of carbon isotopes in $CO_2$ samples collected during the BGS campaign.

Table 1 Summary of all measured parameters and sampling specifications including their minima, maxima and mean values from the survey in 2017 (May-June); Lab = laboratory; b.s. = below surface; a.s. = at surface

| Parameters | Dimension of study area [km] | Grid spacing [m] | N | Sampling procedure/ Analysis | Sampling time [min] | Min | Max | Mean |
|---|---|---|---|---|---|---|---|---|
| $CO_2$ efflux [$g\ m^{-2}d^{-1}$] | 6 × 4 | 25 x 200 | 2,823 | a.s./ In-situ | 1-2 | 0 | 839 | 8.5 |
| $\delta^{13}C\text{-}CO_2$ [δ ‰ vs. VPBD] | Selected sites | Single points | 38 | 1m b.s./ Lab | approx. 10 | -19.2 | -1.2 | -7.9 |
| $T_s$ [°C] | 5.8 × 2.4 | 50/100 × 200 | 858 | 50 cm b.s./ In-situ | approx. 10 | 5.9 | 91.3 | 17.5 |
| $^3He/^4He$ [$R/R_a$] | Selected sites | Single points | 6 | Variable depth (max. 30 cm b.s)/ Lab | 10 | 2.31 | 4.88 | 3.4 |

Table 2 Summary of all measured parameters with minimum, maximum and mean values from the surveys in 2018 (February & April – Area D). The analysis and sampling procedure are given in Table 1. Location of Areas A to F can be seen in Figure 3a. Note that results of Area F are not shown in Fig. 3b, since emissions are mainly background (≤ 22. 4 $g\ m^{-2}\ d^{-1}$) and carbon isotopic values indicate a biogenic origin of $CO_2$. VPBD = Vienna Pee Dee Belemnite

| | Dimension of study area [m] | Grid spacing [m] | CO₂ emissions | | | | Carbon isotopes | | | | Soil temperatures | | | |
|---|---|---|---|---|---|---|---|---|---|---|---|---|---|---|
| | | | N $CO_2$ efflux [$g\ m^{-2}d^{-1}$] | Min | Max | Mean | N $\delta^{13}C\text{-}CO_2$ [δ ‰ vs. VPBD] | Min | Max | Mean | N $T_s$ [°C] | Min | Max | Mean |
| **Area A** | 400 × 300 | 25 - 30 × 90 - 120 | 64 | 0.71 | 1,464.2 | 61.1 | 7 | -18.0 | -2.9 | -5.9 | 10 | 12.2 | 22.1 | 18.1 |
| **Area B** | 300 × 200 | 25 - 30 × 90 - 120 | 40 | 2.12 | 63.8 | 22.7 | 10 | -20.5 | -3.7 | -9.8 | 39 | 13.1 | 52.3 | 27.2 |
| **Area D** | 140 × 260 | 10 × 10 | 480 | 0.16 | 526.2 | 25.2 | 6 | -18.4 | -2.4 | -7.2 | - | - | - | - |
| **Area E** | 770 × 500 | 25 - 30 × 90 - 120 | 131 | 0.49 | 1,285.5 | 44.5 | 23 | -19.8 | -1.6 | -6.5 | 128 | 8.3 | 36 | 18.1 |
| **Area F** | 500 × 350 | 25 - 30 × 90 - 120 | 83 | 1.88 | 22.4 | 8.55 | 8 | -23.2 | -12.7 | -18.4 | 83 | 5.7 | 22.6 | 14.1 |

Table 3 Statistical parameters of the total $CO_2$ efflux dataset obtained by the graphical statistical analysis.

| Population | Proportion [%] | N | Mean $CO_2$ efflux [g m$^{-2}$ d$^{-1}$] | $CO_2$ efflux range [g m$^{-2}$ d$^{-1}$] |
|---|---|---|---|---|
| **Background Population (A)** | 90.2 | 3,203 | 6 | 0.1 – 20.7 |
| **Mixed Population (B)** | 9.7 | 344 | 66.4 | 20.7 – 640.8 |
| **Hydrothermal Population (C)** | 0.1 | 5 | 1,038.1 | 640.8 – 1,464.2 |

Table 4 Summary of CO$_2$ output estimations from the large-scale and small-scale surveys at Los Humeros. For comparison we included values from the study of Peiffer et al. (2018) as well as other volcanic/geothermal systems worldwide.

| Study Area | Area [km²] | Total CO$_2$ output [t d$^{-1}$] | Standard deviation | Reference |
|---|---|---|---|---|
| **Total study area** | 13.6 | 87.3 | 10.2 | This study |
| **Total study area above background threshold** | 0.5 | 26.1 | 0.078 | |
| **Area A** | 0.13 | 1.6 | $8 \times 10^{-5}$ | |
| **Area B** | 0.06 | 1.3 | $9.1 \times 10^{-5}$ | |
| **Area D** | 0.04 | 0.7 | $1.2 \times 10^{-4}$ | |
| **Area E** | 0.43 | 10.6 | $1.5 \times 10^{-4}$ | |
| **Humeros North** | 0.06 | 13.5 | 3.9 | Peiffer et al., 2018 |
| **Humeros South** | 0.012 | 2.66 | 0.3 | |
| **Xalapasco** | 0.0045 | 1.38 | 0.05 | |
| **Volcanic/Geothermal system Kos, Greece** | 0.4 | 74.7[a] | n/a | Daskalopoulou et al., 2019 |
| **Volcanic/geothermal system Monte Amiata, Italy** | 0.22 | 221 | 25 | Frondini et al., 2009 |
| **Volcanic/geothermal system Reykjanes, Iceland** | 0.23 | 13.9 | 1.7 | Fridriksson et al., 2006 |
| **Geothermal system Ohaaki, New Zealand** | 12.7 | 111 | 6.7 | Rissmann et al., 2012 |

[a] Only CO$_2$ efflux values above biogenic threshold are included

*Table 5 Statistical parameters of $\delta^{13}C\text{-}CO_2$ obtained by the GSA method and corresponding $CO_2$ efflux values.*

| Population | Proportion [%] | N | $\delta^{13}C\text{-}CO_2$ range [δ ‰ vs. VPBD] | Mean $\delta^{13}C\text{-}CO_2$ [δ ‰ vs. VPBD] | Associated $CO_2$ efflux range [g m$^{-2}$ d$^{-1}$] | Associated Mean $CO_2$ efflux [g m$^{-2}$ d$^{-1}$] |
|---|---|---|---|---|---|---|
| **Biogenic (A)** | 13 | 12 | -23.2 to -18 | -19.9 | 2.6 – 20.6 | 9 |
| **Mixed (B)** | 46 | 36 | -17.9 to -5.96 | -10.8 | 1.7 – 839 | 49.3 |
| **Hydrothermal (C)** | 40 | 42 | -5.92 to -1.2 | -3.6 | 1.0 – 1,464.2 | 190.1 |

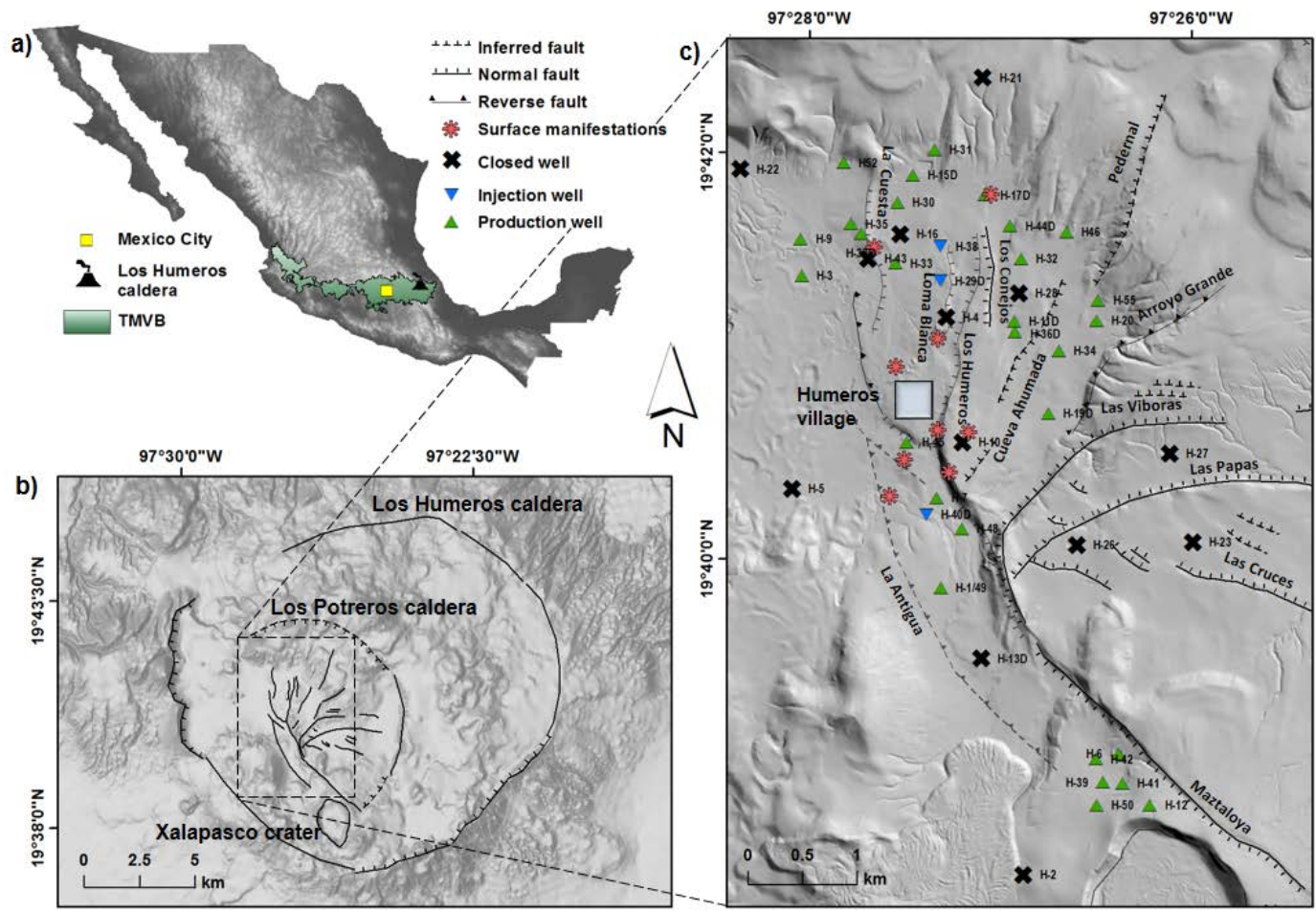

Fig 1. a) Map of Mexico showing the location of the Los Humeros caldera within the Trans Mexican Volcanic Belt (TMVB). b) Digital elevation model (DEM) reproduced with permission of Instituto Nacional de Estadística y Geografía (INEGI) showing mapped and inferred scarps of the Los Humeros and Los Potreros caldera originating from the two main caldera forming events. The youngest evidence for active volcanism inside the caldera is the Xalapasco crater (7.3 ± 0.1 ka) in the south. c) High resolution DEM (1m) from Carrasco-Núñez et al. (2017) of the study area showing all injection (green triangle down) and production wells (blue triangle up). Solid and dashed black lines represent the fault network transferred from Norini et al., 2015. Red asterisks show locations of surface manifestations e.g. advanced argillic alteration, weak steam vents, sulfatara, and warm ground (own observation). Light grey square indicates the location of Humeros village.

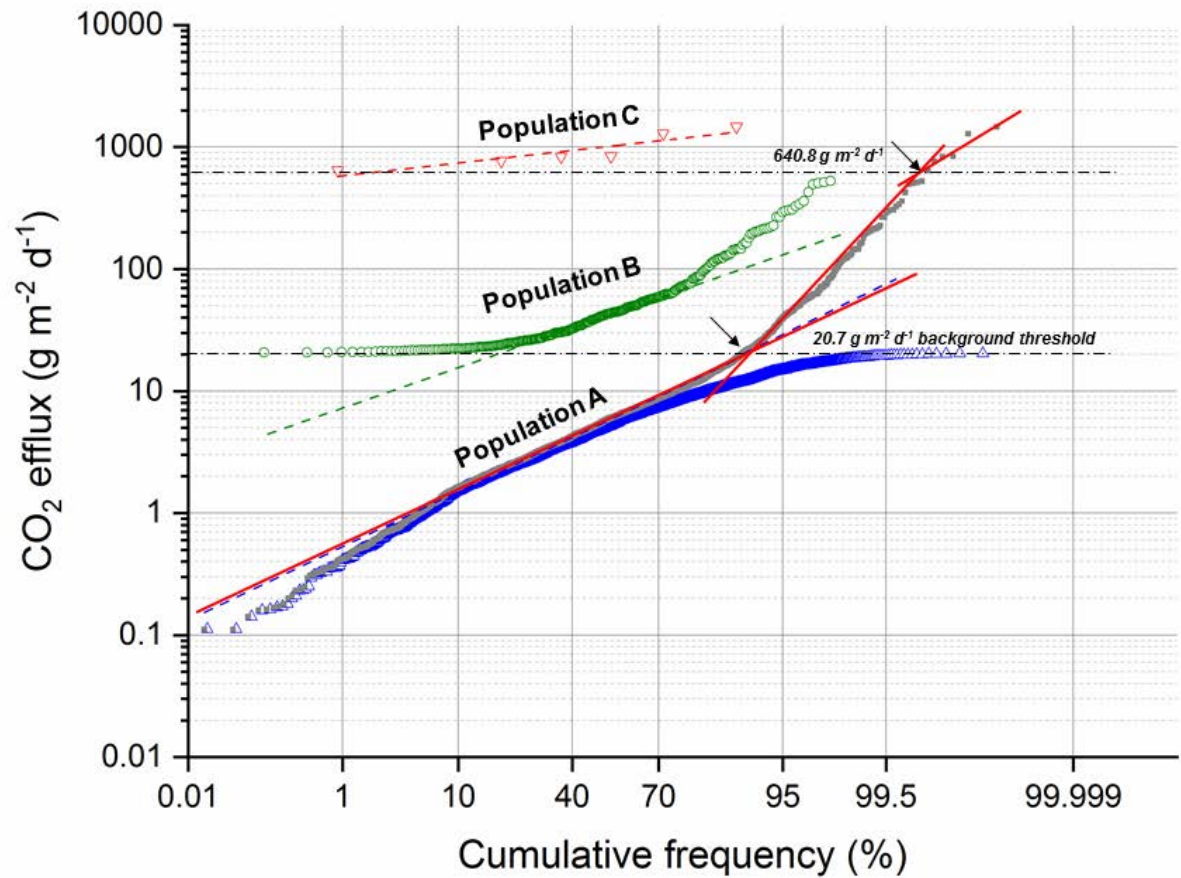

Fig. 2. Probability plot for the entire $CO_2$ efflux dataset. Black arrows point on the inflection points, which divide the dataset in to three populations. Lower dashed black line indicates the threshold value for background $CO_2$ efflux and the upper one indicates the threshold between population B and C.

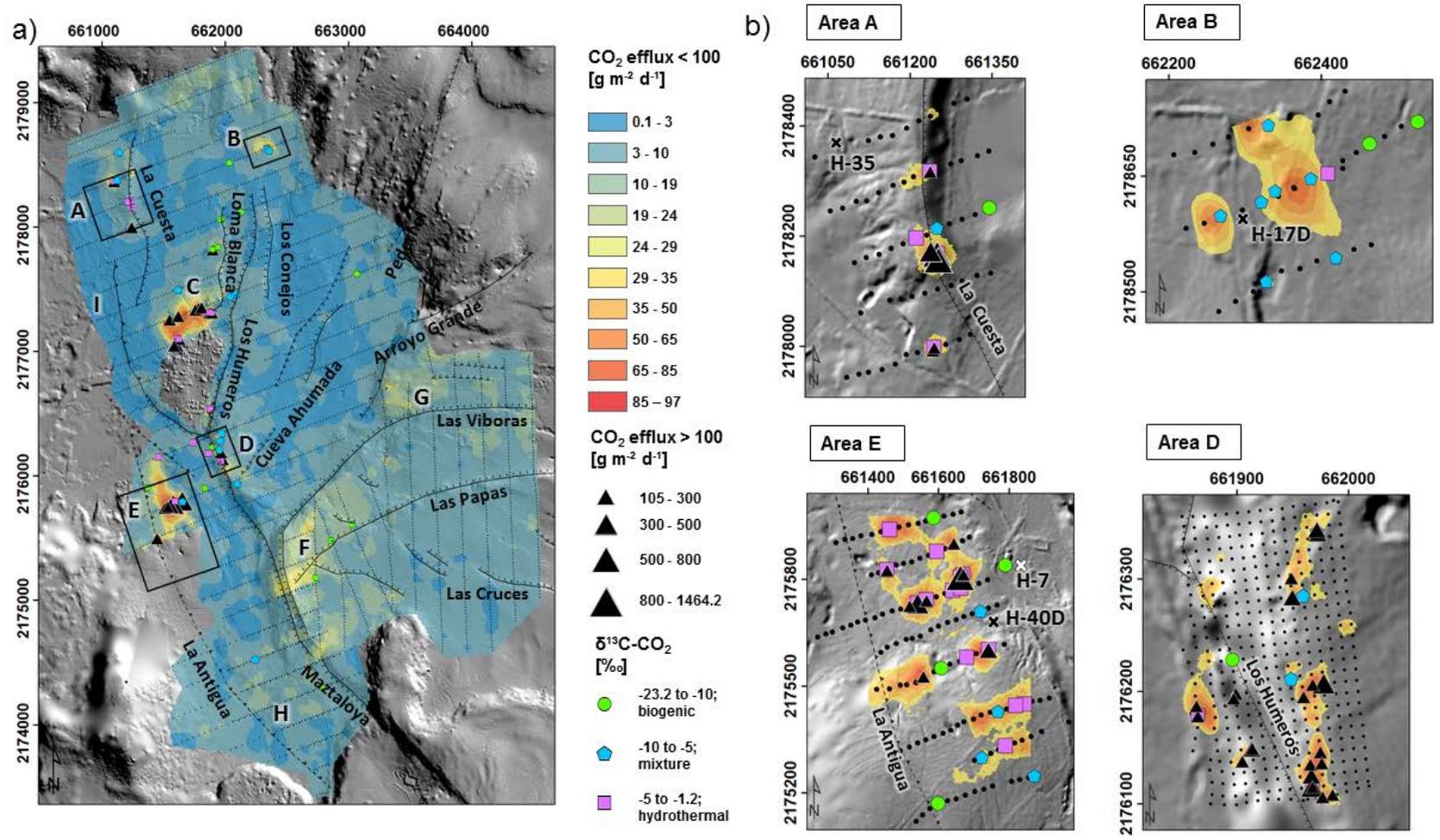

*Fig 3. Results of sequential Gaussian simulation for CO$_2$ efflux a) from 2017 showing the distribution of low, intermediate and elevated degassing sites up to 97 g m$^{-2}$ d$^{-1}$. Black squares (A, B, D, E) show the location and size of the small scale surveys performed in 2018. b) CO$_2$ efflux maps for Area A, B, E, and D. Values lower than 29 g m$^{-2}$ d$^{-1}$ are masked. Labeled black crosses show location of production wells. The white cross in Area E shows an injection well. Graduated black triangles (all maps) illustrate CO$_2$ efflux values > 97 g m$^{-2}$ d$^{-1}$. The classification of carbon isotopic measurements and related symbols applies to all maps. Small black dots represent CO$_2$ efflux sampling sites. Solid and dashed black lines illustrate known and inferred faults. The grey cutout between Area C and D shows Humeros village where no measurements were performed to avoid artificial effects.*

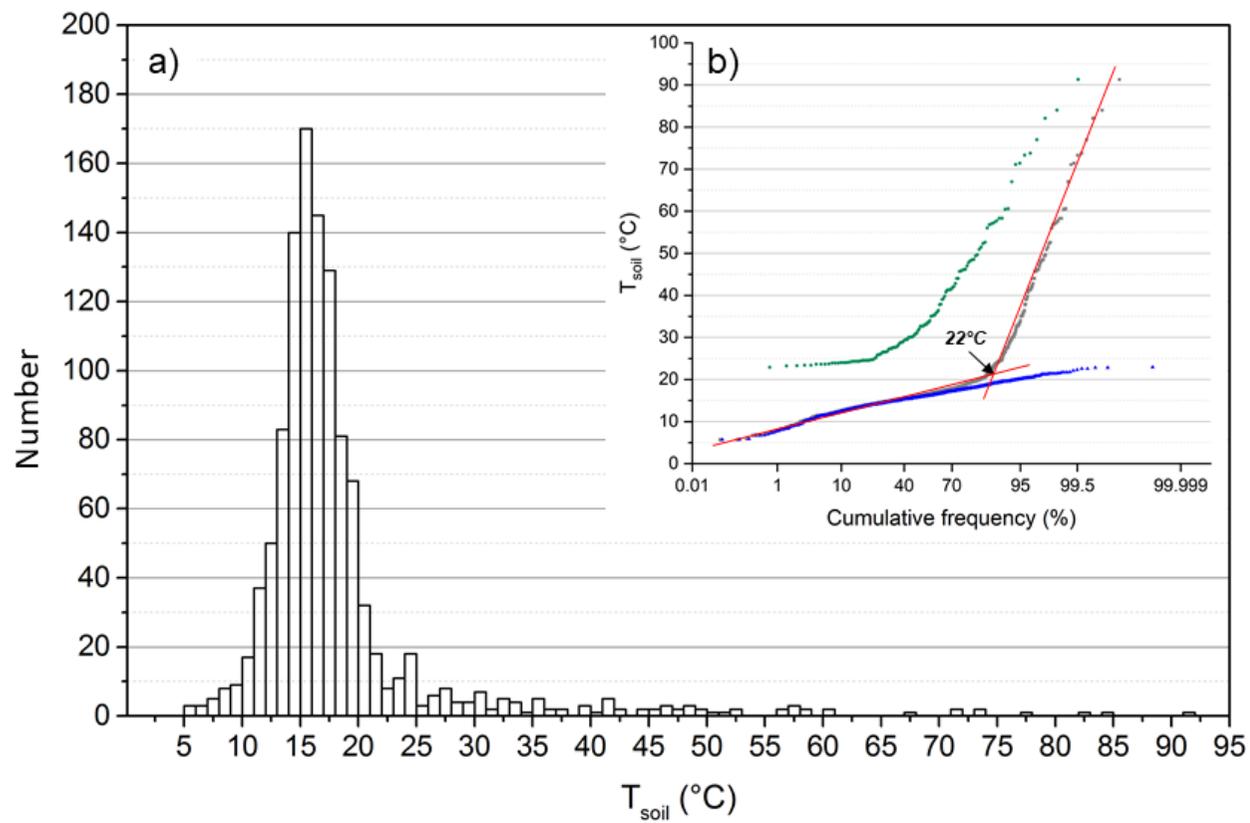

Fig. 4a) Histogram showing a right skewed frequency distribution of soil temperatures. The majority of values (87.9%) range between 5.9 °C and 22 °C. b) Probability plot of soil temperatures indicating a major inflection point at 22°C, which separates the dataset into background and anomalous soil temperatures.

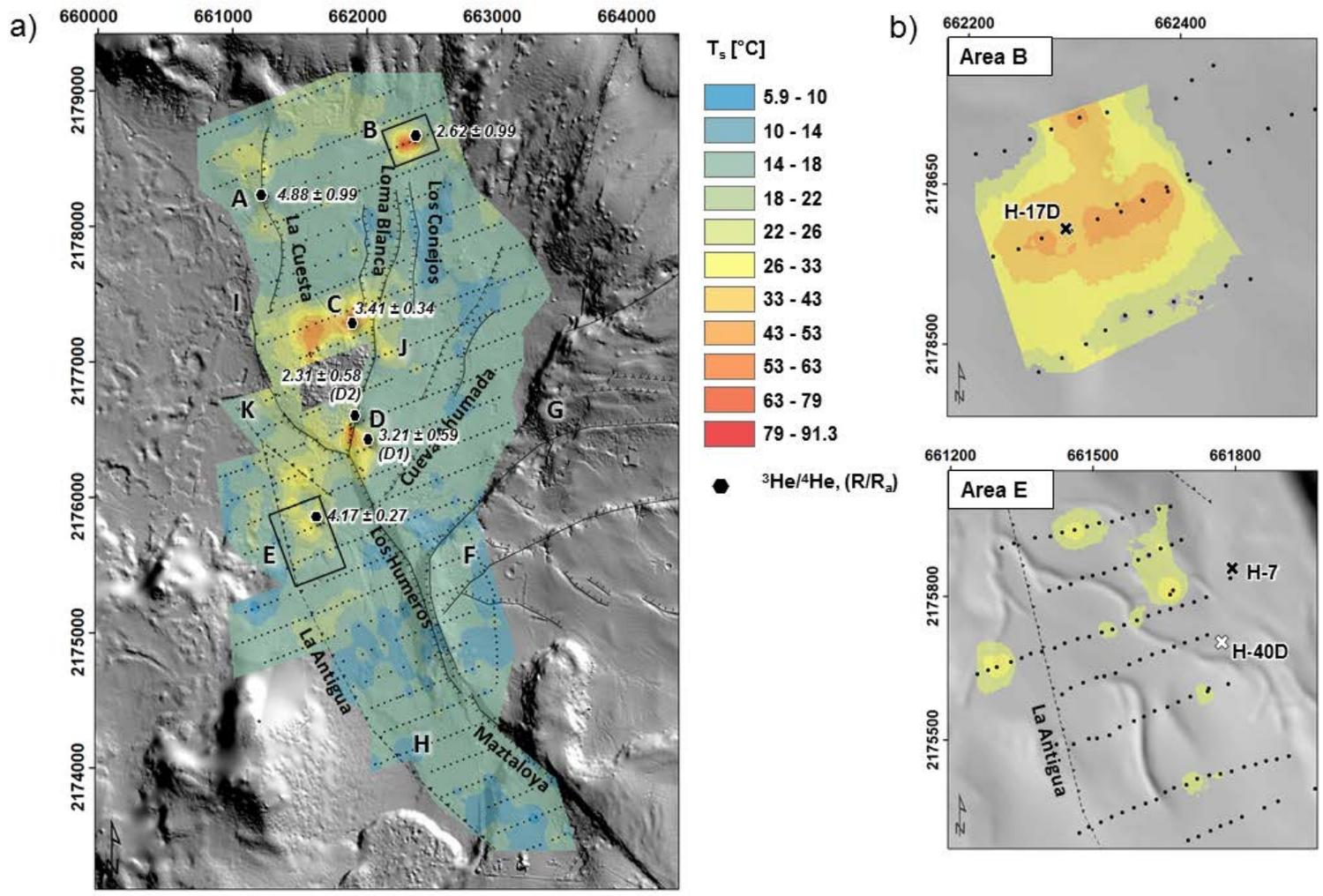

*Fig 5. Results of sequential Gaussian simulation a) for all measured soil temperatures in 2017. Black small dots represent soil temperature sampling sites. Black hexagons illustrate sampling sites and results for air-corrected helium ratios at weak to moderate steam vents. Note that two helium samples were taken in Area D (D1 & D2). Black squares show the location of the small scale surveys for Area B and E. b) Soil temperature maps for Area B, and E. Temperatures below 25°C are masked. Black and white crosses illustrate production and injection wells, respectively.*

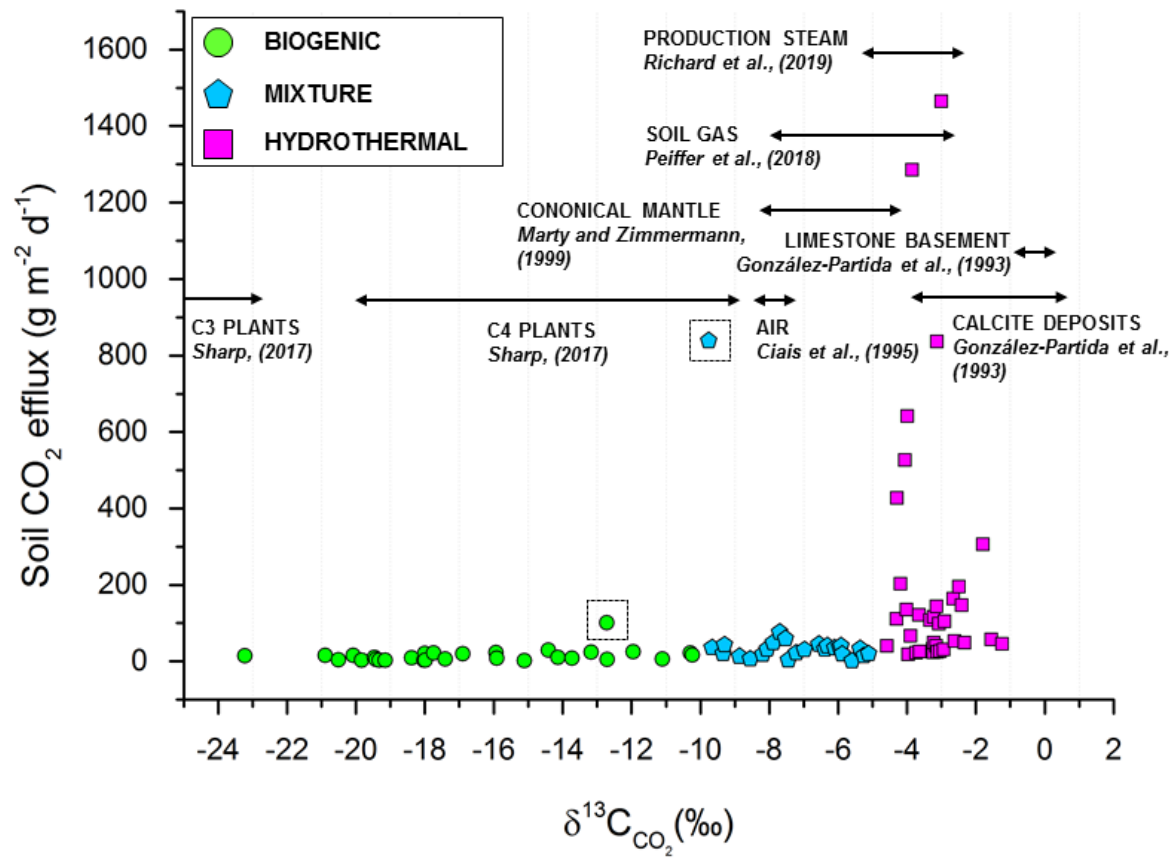

Fig 6. Plot illustrating soil $CO_2$ efflux versus all carbon isotopic composition of soil $CO_2$. Their spatial distribution can be seen in Fig. 3. We illustrate the range of measured carbon isotopes from various studies at Los Humeros for comparison. $\delta^{13}C_{CO2}$ values from increased soil gas emissions and production steam are in accordance with our values from the hydrothermal and mixed groups. Two samples likely show a contamination with air (surrounded by black dashed rectangle). One sample was taken close to injection well H-29 with a $CO_2$ efflux of 100.8 g $m^{-2}$ $d^{-1}$ and a corresponding $\delta^{13}C_{CO2}$ value of -12.7‰, whereas the other sample was taken in Area E with a $CO_2$ efflux of 839 g $m^{-2}$ $d^{-1}$ and a $\delta^{13}C_{CO2}$ value of -9.8‰ (Fig. 3).

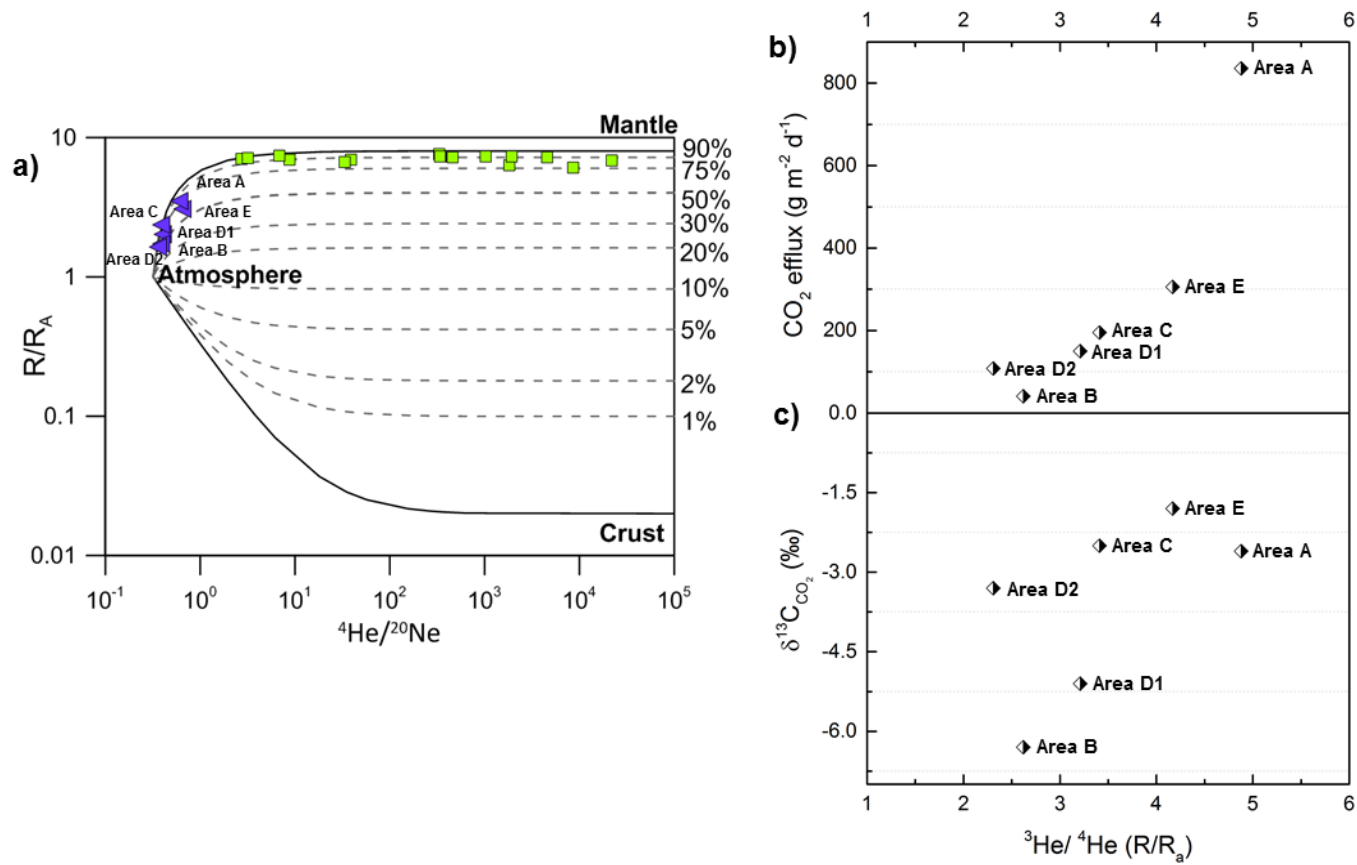

*Figure 7a) Binary plot of R/R$_A$ vs. $^4$He/$^{20}$Ne of samples from steam vents (purple triangles, this study) and geothermal production steam (green squares from Pinti et al., 2017). Dashed lines represent mixing between atmosphere and end-members (crust or mantle) with different percentages of mantle contribution (after Sano and Wakita, 1985). Our results show a clear mixing between atmosphere and up to 65% mantle. b) $^3$He/$^4$He (R/R$_a$) ratios versus CO$_2$ efflux. There is a positive correlation of the two parameters. The highest measured efflux coincides with the highest measured $^3$He/$^4$He ratio. Only Area B shows a slightly higher $^3$He/$^4$He ratio but lower CO$_2$ emissions. c) $^3$He/$^4$He (R/R$_a$) ratios versus carbon isotopic composition of soil CO$_2$. The highest measured $^3$He/$^4$He ratios coincide with $\delta^{13}C_{CO_2}$ values of hydrothermal origin representing the deep geothermal system. Each area is represented by a half-filled diamond. In all diagrams Area A and E show the most evident relation to the superhot geothermal system.*

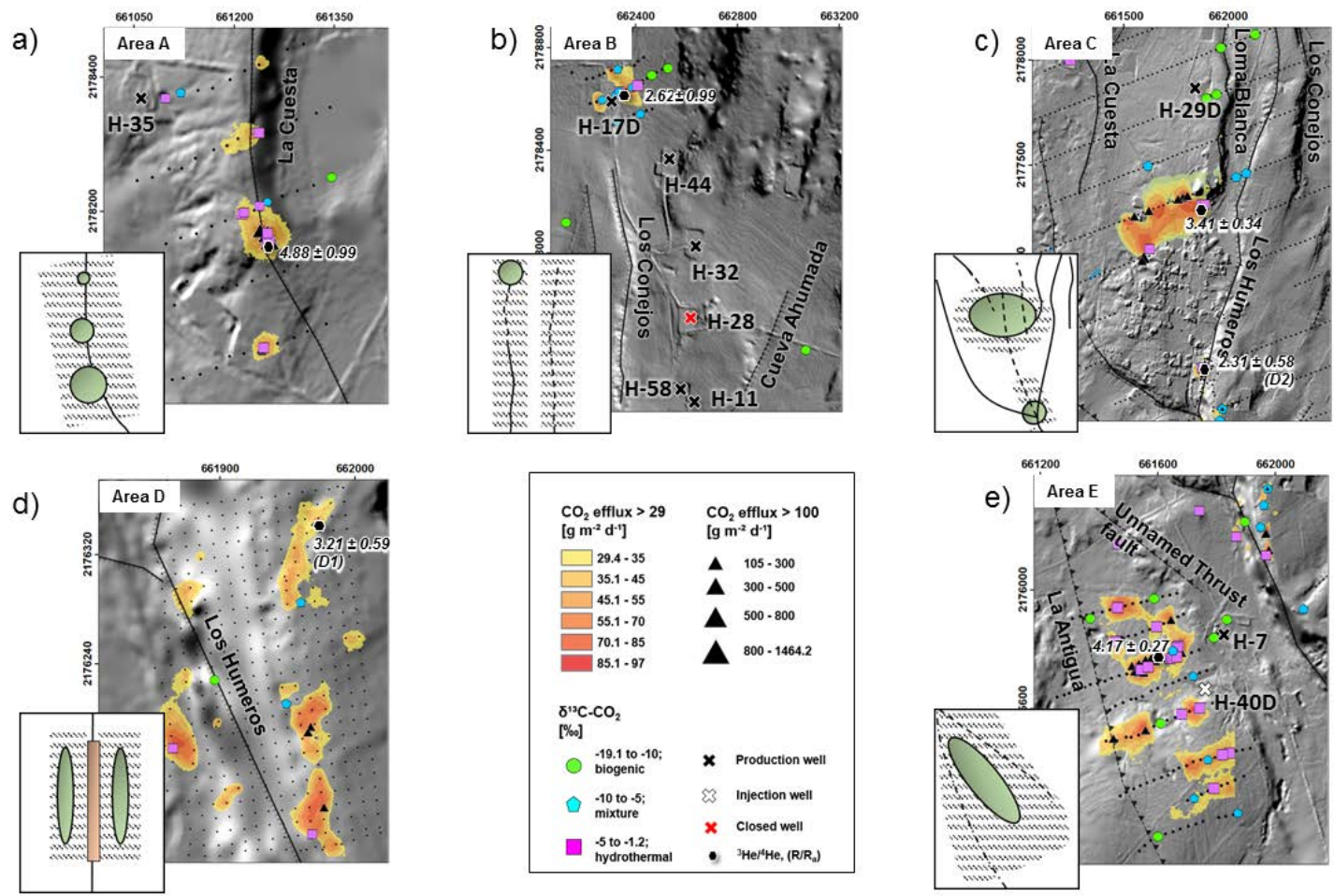

*Figure 8. Detailed view of degassing in Area A, B, C, D, and E. Inset maps provide a schematic (not to scale) structural interpretation based on the catalogue of favorable structural settings for fluid flow by Faulds et al., (2015). Possible dimensions of fault damage zones are outlined (diagonal, dashed, black lines), Green ellipses (upwelling of geothermal fluids) correlate to highest $CO_2$ emissions, impermeable fault core (orange rectangle), known faults (black solid lines), inferred fault (black, dashed lines). a) Shows a typical major normal fault with highest gas emissions at the fault bend. b) Hidden fault or fault continuation of Los Conejos fault. The zoomed out map shows also the location of many production wells drilled along N-S corridor parallel to Los Conejos indicating another structural corridor. c) A horsetail fault termination starting in the south of Humeros village with faults (i.e. La Cuesta, Loma Blanca, and Los Conejos) forming the horse tail. d) Permeable fault damage zone and a less permeable fault core along Los Humeros fault e) Fault intersection or linkage of two fault damage zones favoring increased degassing in Area E.*

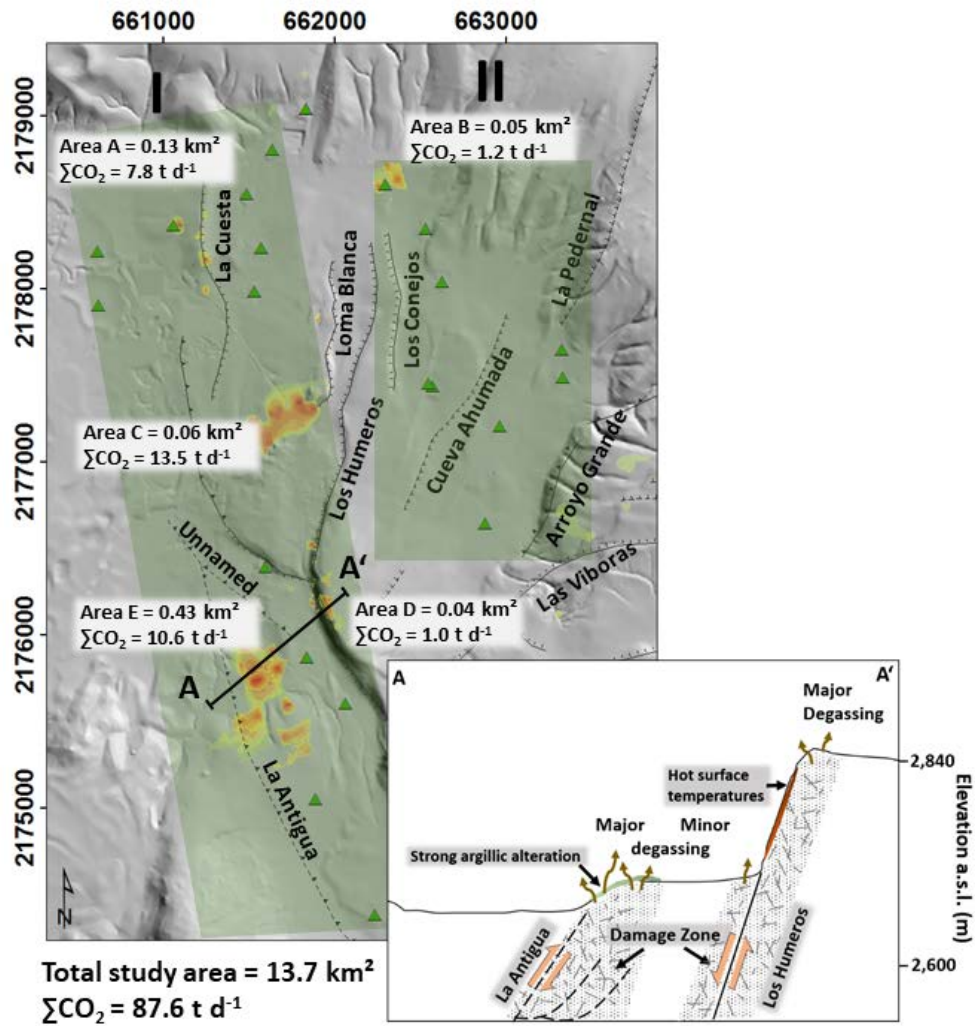

*Figure 9. Simplified map showing two structural corridors (highlighted in green; corridor I = NNW-SSE and corridor II= N-S oriented), which favor hydrothermal fluid flow within the main production zone of the geothermal field. Our interpretation is based on increased $CO_2$ emissions (extracted from Fig. 3) and the location of production wells (green triangles). The total $CO_2$ output and size for each area is shown. Please note that values from Area C are taken from Peiffer et al. (2018) who focused on localized areas of increased degassing rates and a high number of measuring points resulting in a high $CO_2$ emission rate. The cross-section A-A' shows our interpretation of the structural framework in Area D and E. (Legend for $CO_2$ emissions can be found in Fig. 8)*

**Supplementary Material**

All datasets generated and analyzed for this study can be found on the repository of the GFZ Data Service under the following DOI: http://doi.org/10.5880/GFZ.4.6.2020.001

**Figure 1** Experimental and modeled variograms of the CO$_2$ efflux maps for the total study area and the small-scale surveys.

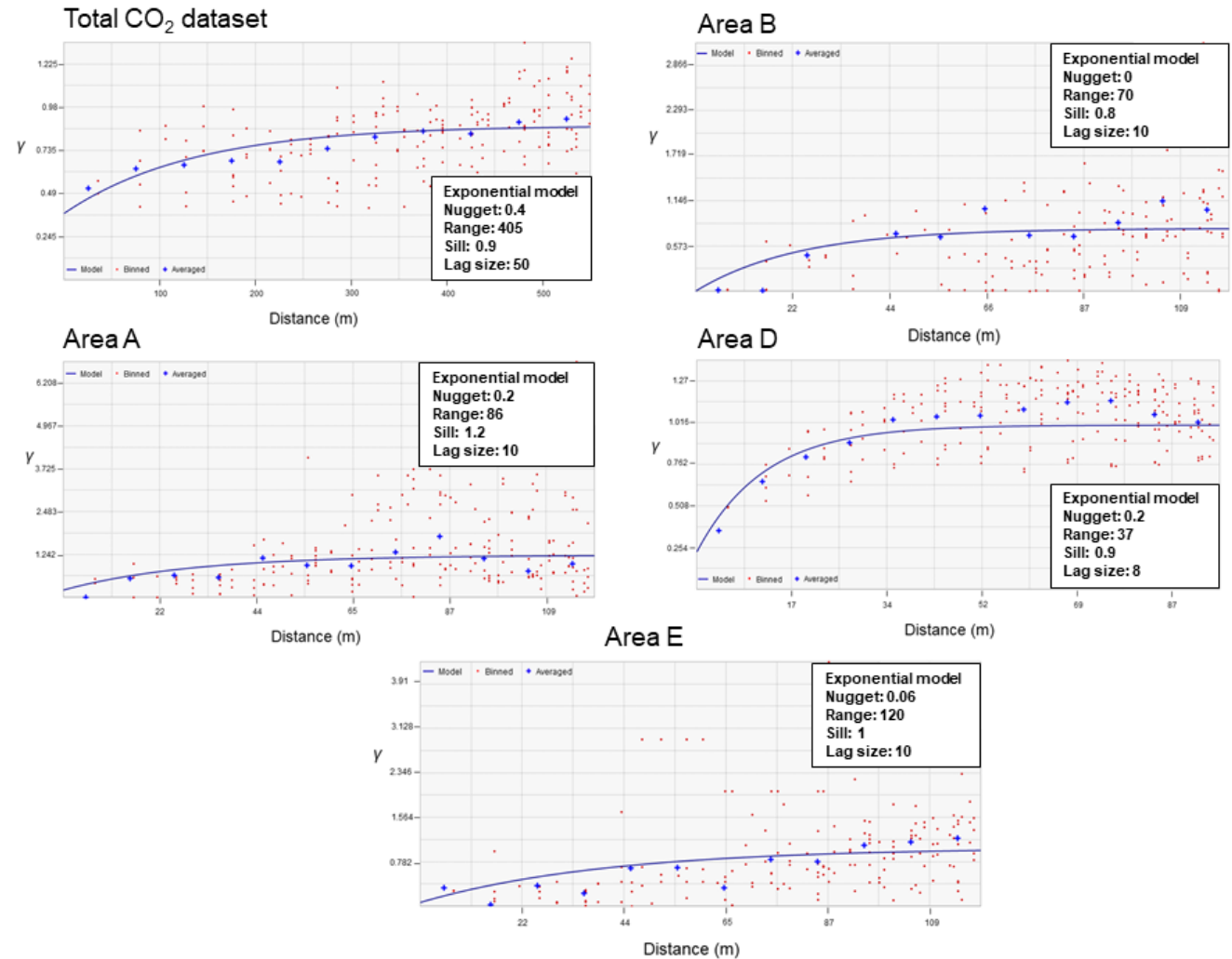

**Figure 2** Experimental and modeled variograms of soil temperature maps for the total study area, Area B, and Area E.

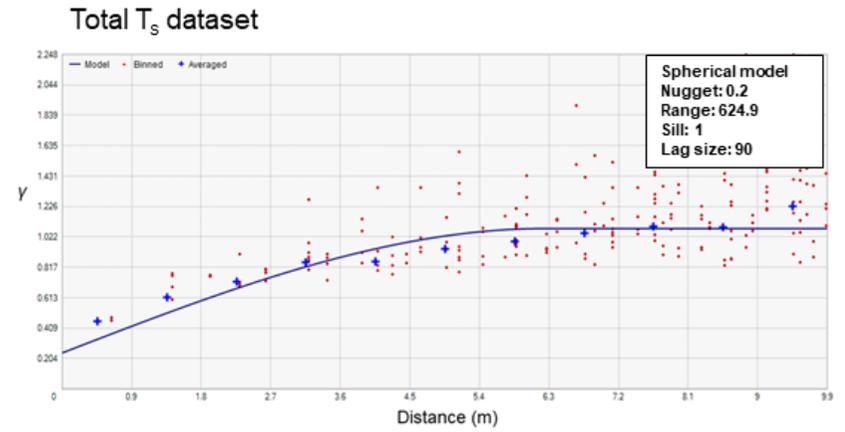

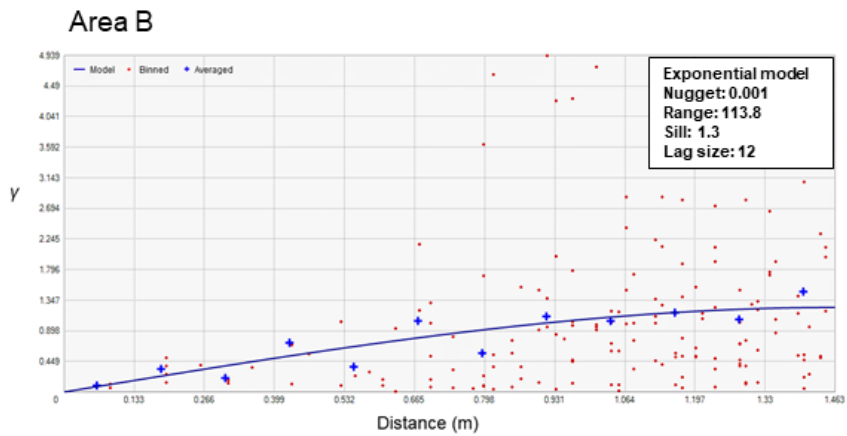

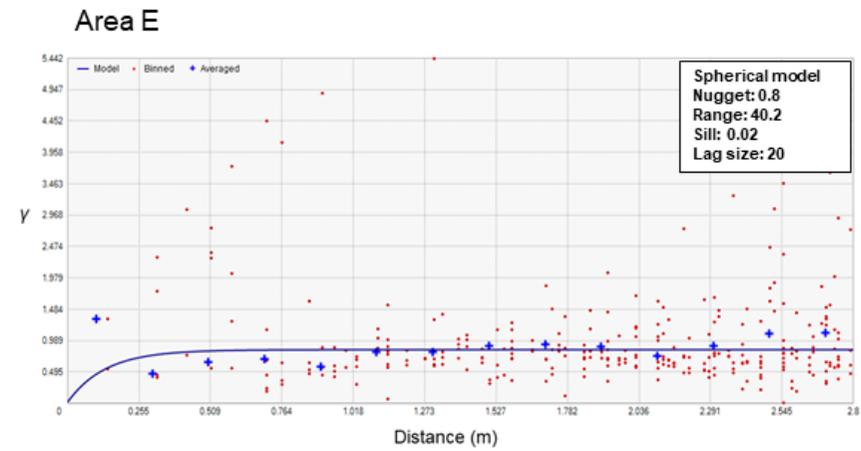

**Figure 3**. Probability plot of carbon isotopes. Based on the identification of two inflection points (black arrows), three populations (biogenic, mixed, and hydrothermal) could be identified.

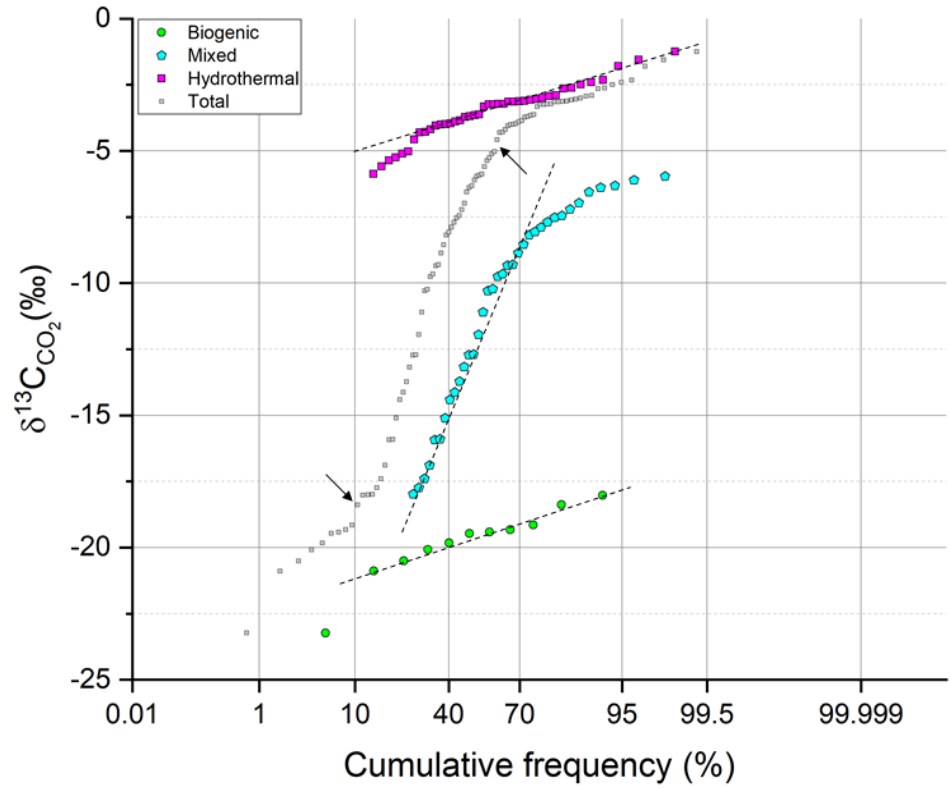